# Reversion of Quantum Walks via interventions on coin space

Mahesh N. Jayakody and Asiri Nanayakkara*
*National Institute of Fundamental Studies,*
*Hanthana Road, Kandy, Sri Lanka*

## Abstract

In this study we show a way of achieving the reverse evolution of $n$-dimensional quantum walks by introducing interventions on the coin degree of freedom during the forward progression of the coin-walker system. Only a single intervention is required to reverse a quantum walker on a line to its initial positon and the number of interventions increases with the dimensionality of the walk. We present an analytical treatment to prove these results. This reversion scheme can be used to generate periodic bounded quantum walks and to control the locations where particle can be found with highest probability. From the point of view of quantum computations and simulations, this scheme could be useful in resetting quantum operations and implementing certain quantum gates.

1. Introduction

Quantum walks (QWs) have become a fruitful testing ground in various areas of science during the past decade. Mainly they contributed to the theoretical and practical improvement in quantum algorithms [1-4] and quantum computing [5, 6]. In addition, QWs have been used to model transport in biological systems [7-9] and other physical phenomena such as Anderson localization [10-14] and topological phases [15, 16].

Unitary evolutions are reversible and hence the quantum walks that employ unitary operators are reversible as well. This obviousness has kept the scientific community away from launching new studies related to reverse evolution of quantum walks. From the point of view of quantum computation and simulations, reversion of quantum walks could be useful in resetting certain quantum operations. Typically, to achieve the reverse evolution of a given quantum walk we need to repeatedly introduce the corresponding adjoint operator into the coin-walker system by replacing the unitary operator which governs the forward progression of the quantum walk. In this paper we present one possible route which can be used to reverse the position state and the coin state of $n$-dimensional QWs without using reversible nature of unitary operators as mentioned above. We introduce limited interventions on the coin degree of freedom during the progression of the walk. Number of interventions depends on the dimensionality. For example, only a single intervention is required for one dimensional quantum walks (1DQWs). Position space is not interfered in anyway. Significantly, this scheme can pull back the quantum walker to its initial position and with the right selection of temporal parameters of this scheme the quantum walk can be made periodic suitably. Such a periodic walk may have favorable consequences in developing quantum gates analogues to classical binary gates. Furthermore, our approach can be modified to induce localization effects in quantum walks provided that the walker starts to move from a single position. The paper is organized in the following way. In section 2 we develop our scheme for 1DQWs and present some important properties of it. Generalization of our results for $n$-dimensional QWs is given in section 3.

## 2. An intervention on coin degree of freedom

First let us recall the standard model of the 1DQW which comprises a two-state coin and a walker. Evolution of the coin-walker system is governed by a unitary operator $U$ defined on a tensor product of two Hilbert spaces, $H_c \otimes H_x$ which are spanned by the coin basis $\{|c\rangle\}_{c \in \{0,1\}}$ and the position basis $\{|x\rangle\}_{x \in \mathbb{Z}}$. Single-step progression of the system is a sequential process in which the coin is tossed at first (transforming the coin state) and then the walker is moved either to the left or right conditional upon the outcome of the coin. Unitary operator that corresponds to a single-step evolution of the system is given by

$$U = S.\, \mathbb{I} \otimes C \tag{1}$$

where $S$ and $C$ are Shift and Coin operators respectively. In this study we stick ourselves to the conventional shift operator $S$ given in (2).

$$S = \sum_{x,c=0,1} |c\rangle\langle c| \otimes |x + (-1)^{c+1}\rangle\langle x| \tag{2}$$

General form of the coin operator [17] that governs the 1DQW can be written as

$$C = \begin{pmatrix} \cos(\theta) & e^{i\phi_1}\sin(\theta) \\ e^{i\phi_2}\sin(\theta) & -e^{i(\phi_1+\phi_2)}\cos(\theta) \end{pmatrix} \tag{3}$$

where $\theta \in [0, 2\pi)$, and $\phi_1, \phi_2 \in [0, \pi)$. Product and Bell states are used as initial coin states. The general state vector of the coin-walker system is written as

$$|\psi(x,t)\rangle = \sum_x a_x(t)|0\rangle|x\rangle + b_x(t)|1\rangle|x\rangle \tag{4}$$

where $\sum_x |a_x(t)|^2 + |b_x(t)|^2 = 1$ and $t$ is the time step. Let the initial state of the system be $|\psi_0\rangle$. Thus the final state of the system after $t$ steps is $|\psi(x,t)\rangle = U^t |\psi_0\rangle$.

As stated earlier, we introduce a single intervention only on the coin degree of freedom by applying a new coin operator at a specific time step during the evolution of the coin-walker system. Let us define an operator $G$ of the form

$$G = \begin{pmatrix} 0 & e^{i\phi_1} \\ -e^{i\phi_2} & 0 \end{pmatrix} \tag{5}$$

Note that $GG^\dagger = G^\dagger G = \mathbb{I}$. Thus $G$ represents a coin operator. Evolution of the coin-walker system under $G$ can be expressed by the unitary operator $V$ written as $V = S.\, \mathbb{I} \otimes G$ where $S$ is the shift operator given in (2). Combining $U$ and $V$ accordingly we model a scheme of quantum walk in which a single intervention on coin space is introduced at a specific time step as;

$$|\psi(x,t)\rangle = U^{t_2} V U^{t_1} |\psi_0\rangle \tag{6}$$

where $t = t_1 + t_2 + 1$.

In general, final state after $t = t_1 + t_2 + 1$ can be written as

$$|\psi(x,t)\rangle = \left(-e^{i(\phi_1+\phi_2)}\right)^{t_2+1}(\mathbb{I}\otimes D)(U^\dagger)^{t_2+1}(U)^{t_1}|\psi_0\rangle \tag{7}$$

where $D = C^\dagger G$. The derivation of (7) is given in the Appendix. By changing $t_1$ and $t_2$ appropriately we can control the reverse and forward evolution of the walker at will. Note that there is no any physical meaning of the overall phase factor in (7) as it does not change the expected values of the Hermitian operators in either position or coin space. In the perspective of density operators of pure states, we can simply ignore the presence of the overall phase factor.

Suppose we choose $t_1 = l$ and $t_2 = l - 1$ where $l$ is an arbitrary number of steps. Final state after $t = 2l$ time steps is given by

$$|\psi(x, 2l)\rangle = U^{(l-1)}VU^l|\psi_0\rangle = \left(-e^{i(\phi_1+\phi_2)}\right)^l (\mathbb{I}\otimes D)|\psi_0\rangle \tag{8}$$

According to (8) position space remains unchanged after $t = 2l$ time steps. That is, after $t = 2l$ time steps particle returns to its initial positon irrespective of the initial coin state. Notice that the coin state after $t = 2l$ is different than that of the initial coin state.

Now consider a routine in which we apply operator $U$ for $t = l$ time steps on the initial state. Resultant state is $|\psi_l\rangle = (U)^l|\psi_0\rangle$. Next operator $V$ is applied for a single step to obtain $|\psi_{l+1}\rangle = V|\psi_l\rangle$. Afterwards we allow the resultant state ($|\psi_{l+1}\rangle$) to evolve continuously under $U$ and introduce $V$ into the quantum walk with $2l - 1$ time gaps. This can be expressed as:

$$|\psi(x, (2m+1)l)\rangle = \left(U^{(2l-1)}V\right)^m U^l |\psi_0\rangle \tag{9}$$

where $m \in \mathbb{N}$. We can rewrite (9) in the following way

$$\begin{aligned}\left(U^{(2l-1)}V\right)^m U^l|\psi_0\rangle &= \cdots U^{(2l-1)}VU^{(2l-1)}VU^{(2l-1)}VU^l|\psi_0\rangle \\ &= \cdots [U^{(l-1)}VU^l][U^{(l-1)}VU^l]|\psi_0\rangle\end{aligned} \tag{10}$$

By using (8) and the property $DD = \mathbb{I}$ we can easily show that (10) represents a periodic quantum motion. It is worth noticing that the periodicity of the coin state is two times greater than that of the positon state. This is illustrated in Figure 1. The periodicity of the whole coin-walker system is equal to that of the coin state. These asynchronous periods in positon and coin state could be utilized to implement a quantum analogue to classical frequency divider that is used in classical computers as a binary divider. Typically, a frequency divider is triggered by an external clock pulse. In the quantum analogue it is equivalent to the event of introducing the operator $V$ with $2l - 1$ time gaps. The other significant feature of our periodic motion is that the walker turns back at the initial position by itself without any external triggering.

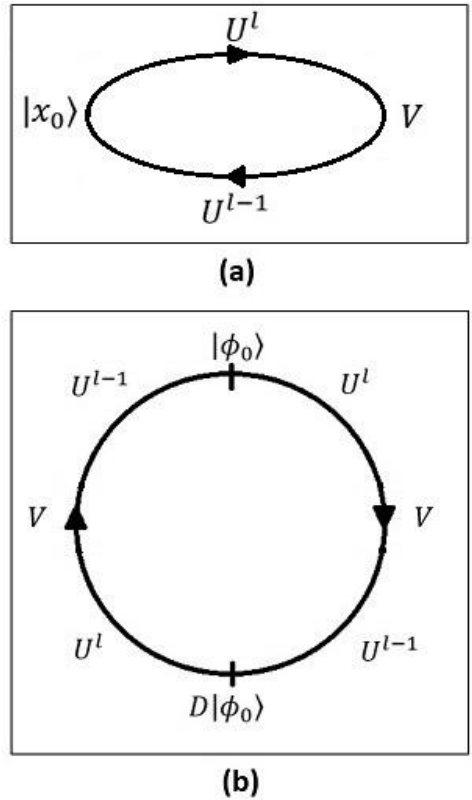

FIG:1 Single full cycle of position state (a) and coin state (b) of the periodic quantum motion

We have shown that the routine given in (10) can back pedal the position state into its initial state periodically. This is true for both localized and non-localized initial positon states. To exhibit a finite periodicity in the positon state, quantum walker must spread over in a bounded region on the line during a period of time. Therefore, the routine defined in (10) can be used to generate bounded quantum walks. An interesting thing can be observed when the routine starts from a localized positon state. As a result of periodic motion we obtain the localized position state repeatedly. Hence, one can use this scenario to generate periodic localization effects in quantum walks.

Another feature of the quantum walk with a single intervention can be described as follows. Consider a quantum walk in which the walker starts from a single position carrying a product or Bell state as the initial coin state. Walker is allowed to evolve for a fixed number of time steps. The positon distribution of such a walk always has a peak and it appears close to left or right end of the distribution. This means that the probability of finding the walker at one end of the position axis is high. By introducing a single intervention at a proper time step during the evolution, we can move this probability peak along the positon axis. In other words, for a given initial coin state (product or Bell state) and a fixed number of time steps, we can control the location in which the walker can be found with the highest probability by introducing a single intervention at a proper time step. This is illustrated in Figure 2. The usual ballistic spread of the Hadamard Walk for the initial state $|\psi_{in}\rangle = |1\rangle_c \otimes |0\rangle_x$ and 100 time steps is given in Fig 2 (d). Under this initial state, probability of finding the walker in the

negative positon region is comparatively low. However, by introducing a single intervention at a proper time step we can increase that probability up to a substantial level (see Fig 2 (a)). When the intervention is introduced in the midst of the evolution, walker returns to the initial position as predicted in (7). This is shown in Fig 2 (b). Furthermore observe that in Fig 2 (d) the Hadamard Walk for 100 time steps is bounded to the region $-75 < x < 75$. However, by introducing a single intervention at a proper time step we can define different boundaries for the 'Hadamard Walk for 100 time steps' as shown in Fig 2 (a) and (b).

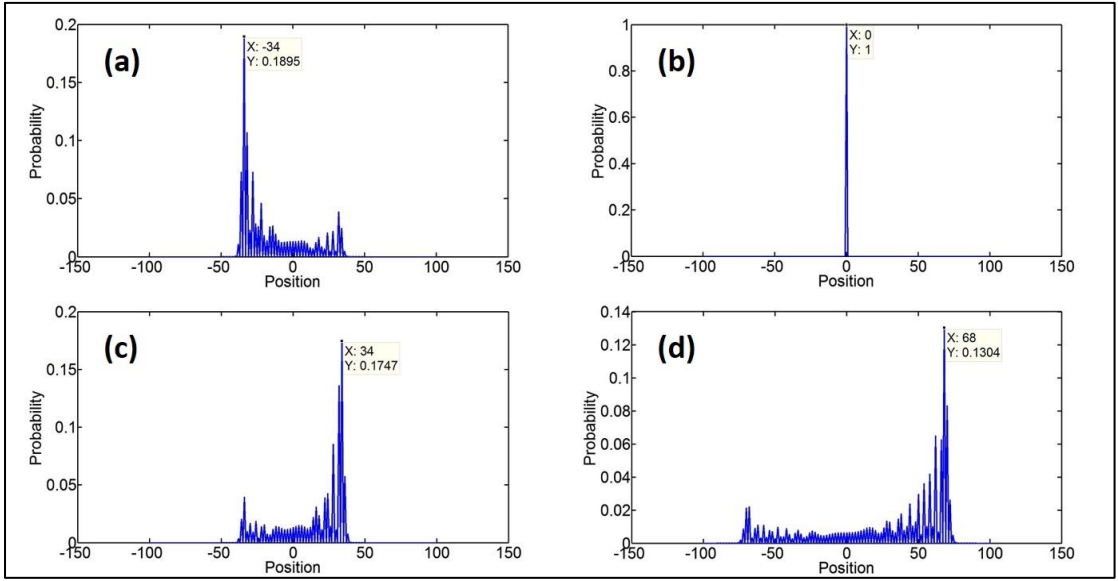

FIG:2 Probability distributions for Hardmand walk with and without interventions. Initial state is $|\psi_{in}\rangle = |1\rangle_c \otimes |0\rangle_x$ and $t = 100$ time steps (a) intervention at $t = 26$ (b) intervention at $t = 51$ (c) intervention at $t = 76$ (d) No intervention.

3. Interventions on higher dimensional coin spaces

Reversion idea in 1DQW can be generalized to higher dimensions very easily. First let us derive the useful mathematical tools and afterwards pay attention on the generalization of the above result. For the sake of simplicity let us conduct our analysis in k-basis. Unitary operator that governs a one dimensional quantum walk in momentum space can be written as

$$U_k = (\mathbb{I} \otimes C_k) \qquad (11)$$

where $C_k = (e^{-ik}P_R + e^{ik}P_L)C$. Note that $P_R$ and $P_L$ are two orthogonal projectors on coin Hilbert Space and $C$ is the usual coin operator. This representation allows us to view an operation performed on an arbitrary state as an operation performed only on the coin state of that arbitrary state. A unitary operator for higher dimensional quantum walks can be deduced from (9). Let $C_k^{(n)}$ be an unitary operator for the quantum walk on n-dimensional space where $n > 1$. Suppose $(e^{i\omega_1}, ..., e^{i\omega_{2^n}})$ and $\{|\phi_1\rangle, ..., |\phi_{2^n}\rangle\}$ are the non-degenerate eigenvalues and corresponding eigenvectors of $C_k^{(n)}$ respectively. Then we can write the spectral decomposition of $C_k^{(n)}$ as:

$$C_k^{(n)} = \sum_{k=1}^{2^n} e^{i\omega_k} |\phi_k\rangle\langle\phi_k| \tag{12}$$

Let $t$ be an arbitrary number of time steps. Then we can write the $t$ steps evolution of the operator $C_k^{(n)}$ as follows:

$$\left(C_k^{(n)}\right)^t = \sum_{k=1}^{2^n} e^{i\omega_k t} |\phi_k\rangle\langle\phi_k| \tag{13}$$

Define a unitary operator of the form:

$$G_k = \sum_{j=2}^{2^n} |\phi_j\rangle\langle\phi_{j-1}| + |\phi_1\rangle\langle\phi_{2^n}| \tag{14}$$

The expression for the operator $\left(C_k^{(n)}\right)^t G_k$ is obtained by multiplying (13) and (14).

$$\left(C_k^{(n)}\right)^t G_k = \sum_{k=2}^{2^n} e^{i\omega_k t} |\phi_k\rangle\langle\phi_{k-1}| + e^{i\omega_1 t} |\phi_1\rangle\langle\phi_{2^n}| \tag{15}$$

Then the $m^{th}$ power of $\left(C_k^{(n)}\right)^t G_k$ where $1 < m < 2^n$ can be written as:

$$\left(\left(C_k^{(n)}\right)^t G_k\right)^m = \sum_{k=m+1}^{2^n} \left(\prod_{l=k-(m-1)}^{k} e^{i\omega_l t}\right) |\phi_k\rangle\langle\phi_{k-m}| + $$
$$\sum_{j=1}^{m-1} \left(\prod_{u=1}^{j} e^{i\omega_u t}\right) \left(\prod_{v=0}^{m-j-1} e^{i\omega_{(2^n-v)} t}\right) |\phi_j\rangle\langle\phi_{2^n-(m-j)}| + \left(\prod_{u=1}^{m} e^{i\omega_u t}\right) |\phi_m\rangle\langle\phi_{2^n}| \tag{16}$$

From the method of induction it can be proved that (16) is valid for each $m$ where $1 < m < 2^n$. By substituting $m = 2^n - 1$ we get:

$$\left(\left(C_k^{(n)}\right)^t G_k\right)^{2^n-1} = \left(\prod_{l=2}^{2^n} e^{i\omega_l t}\right) |\phi_{2^n}\rangle\langle\phi_1| + \left(\prod_{l=1}^{2^n-1} e^{i\omega_l t}\right) |\phi_{2^n-1}\rangle\langle\phi_{2^n}| + $$
$$\sum_{j=1}^{2^n-2} \left(\prod_{u=1}^{j} e^{i\omega_u t}\right) \left(\prod_{v=0}^{2^n-j-2} e^{i\omega_{(2^n-v)} t}\right) |\phi_j\rangle\langle\phi_{j+1}| \tag{17}$$

Using (12) and (17), an expression for the operator $\left(\left(C_k^{(n)}\right)^t G_k\right)^{2^n-1} \left(C_k^{(n)}\right)^t$ can be written as follows:

$$\left(\left(C_k^{(n)}\right)^t G_k\right)^{2^n-1} \left(C_k^{(n)}\right)^t = \left(\prod_{l=1}^{2^n} e^{i\omega_l t}\right) |\phi_{2^n}\rangle\langle\phi_1| + \left(\prod_{l=1}^{2^n} e^{i\omega_l t}\right) |\phi_{2^n-1}\rangle\langle\phi_{2^n}| + $$
$$\sum_{j=1}^{2^n-1} \left(\prod_{u=1}^{j-1} e^{i\omega_u t}\right) \left(\prod_{v=0}^{2^n-j-1} e^{i\omega_{(2^n-v)} t}\right) e^{i\omega_j t} |\phi_j\rangle\langle\phi_{j+1}| \tag{18}$$

Note that for each $j$ the following relationship holds:

$$\left(\prod_{u=1}^{j-1} e^{i\omega_u t}\right)\left(\prod_{v=0}^{2^n-j-1} e^{i\omega_{(2^n-v)}t}\right) e^{i\omega_j t} = \left(\prod_{l=1}^{2^n} e^{i\omega_l t}\right) \quad (19)$$

Thus we can write;

$$\left((C_k^{(n)})^t G_k\right)^{2^n-1} \left(C_k^{(n)}\right)^t = \left(\prod_{l=1}^{2^n} e^{i\omega_l t}\right)\left(\sum_{j=2}^{2^n} |\phi_{j-1}\rangle\langle\phi_j| + |\phi_{2^n}\rangle\langle\phi_1|\right) = \left(\prod_{l=1}^{2^n} e^{i\omega_l t}\right) G_k^\dagger \quad (20)$$

The reversion in multidimensional space can be achieved by performing following operations on a $n$-dimensional quantum walk that begins with the state $|k_0\rangle \otimes |\psi_0\rangle$.

**Operation 01:**
First apply the operator $G_k$ on the initial state to obtain $|k_0\rangle \otimes |\psi_1\rangle = (\mathbb{I} \otimes G_k)|k_0\rangle \otimes |\psi_0\rangle$

**Operation 02:**
Next $C_k^{(n)}$ is applied on the resultant state for any $l$ arbitrary number of steps to obtain

$$|k_0\rangle \otimes |\psi_{l+1}\rangle = \left(\mathbb{I} \otimes \left(C_k^{(n)}\right)^l\right)(\mathbb{I} \otimes G_k)|k_0\rangle \otimes |\psi_0\rangle$$

**Operation 03:**
$G_k$ is applied again. $|k_0\rangle \otimes |\psi_{1+2}\rangle = (\mathbb{I} \otimes G_k)\left(\mathbb{I} \otimes \left(C_k^{(n)}\right)^l\right)(\mathbb{I} \otimes G_k)|k_0\rangle \otimes |\psi_0\rangle$

**Operation 04:**
$C_k^{(n)}$ is applied for any $l$ arbitrary number of steps to get

$$|k_0\rangle \otimes |\psi_{2l+2}\rangle = \left(\mathbb{I} \otimes \left(C_k^{(n)}\right)^l\right)(\mathbb{I} \otimes G_k)\left(\mathbb{I} \otimes \left(C_k^{(n)}\right)^l\right)(\mathbb{I} \otimes G_k)|k_0\rangle \otimes |\psi_0\rangle$$

**Operation 05:**
Repeat the operations 03 and 04 to obtain the state

$$|k_0\rangle \otimes |\psi_{(l+1)(m+1)}\rangle = \left((C_k^{(n)})^t G_k\right)^m \left(\mathbb{I} \otimes \left(C_k^{(n)}\right)^l\right)(\mathbb{I} \otimes G_k)|k_0\rangle \otimes |\psi_0\rangle$$

Using (20) final state of the walker after $t = 2^n(l+1)$ (i.e. when $m = 2^n - 1$) time steps can be written as:

$$|k_0\rangle \otimes |\psi_{2^n(l+1)}\rangle = \left(\prod_{l=1}^{2^n} e^{i\omega_l l}\right)|k_0\rangle \otimes |\psi_0\rangle \quad (21)$$

At the end of $t = 2^n(l+1)$ time steps, the above configuration yields the initial state with a phase factor. This guarantees that the walker returns to the initial position state after $t = 2^n(l+1)$ time steps with unit probability.

## 4. Conclusion

In this paper we present a reverse mechanism for $n$-dimensional quantum walks without using the reversible nature of the unitary operators. We introduce limited interventions on the coin degree of freedom during the quantum evolution and pedal the walker back to its initial position. A forced periodic bounded motion, localization effects and a mechanism to control the locations where particle can be found with highest probability can be generated from this reversion scheme. From the point of view of quantum computation and simulations, this quantum scheme could be useful in resetting quantum operations and implementing certain quantum gates.

## Appendix

The derivation of the equation (07) is given below. Observe that the following results hold for the operators given in (02), (03) and (05).

Result 01: $(\mathbb{I} \otimes G^\dagger)(\mathbb{I} \otimes C)(\mathbb{I} \otimes G^\dagger) = (\mathbb{I} \otimes C^\dagger)$

Result 02: $(\mathbb{I} \otimes G^\dagger)S(\mathbb{I} \otimes G) = S^\dagger$ and $(\mathbb{I} \otimes G)S(\mathbb{I} \otimes G^\dagger) = S^\dagger$

Result 03: $G^\dagger G^\dagger = -e^{-i(\phi_1+\phi_2)}\mathbb{I}$ and $GG = -e^{i(\phi_1+\phi_2)}\mathbb{I}$

$$U^{t_2} V U^{t_1} |\psi_0\rangle = U^{t_2} V |\psi_{t_1}\rangle$$

$$U^{t_2} V |\psi_{t_1}\rangle = \left(\prod_{t_2} S(\mathbb{I} \otimes C)\right) S(\mathbb{I} \otimes G) |\psi_{t_1}\rangle$$

$$= \left[\prod_{t_2} \left(-e^{i(\phi_1+\phi_2)}\right)(\mathbb{I} \otimes G^\dagger)(\mathbb{I} \otimes G^\dagger) S(\mathbb{I} \otimes G)(\mathbb{I} \otimes G^\dagger)(\mathbb{I} \otimes C)\right] S(\mathbb{I} \otimes G) |\psi_{t_1}\rangle$$

$$= \left[\prod_{t_2} \left(-e^{i(\phi_1+\phi_2)}\right)(\mathbb{I} \otimes G^\dagger) S^\dagger (\mathbb{I} \otimes G^\dagger)(\mathbb{I} \otimes C)\right]\left(-e^{i(\phi_1+\phi_2)}\right)(\mathbb{I} \otimes G^\dagger)(\mathbb{I} \otimes G^\dagger) S(\mathbb{I} \otimes G) |\psi_{t_1}\rangle$$

$$= \left(-e^{i(\phi_1+\phi_2)}\right)^{t_2+1} (\mathbb{I} \otimes C^\dagger)(\mathbb{I} \otimes G)(\mathbb{I} \otimes G^\dagger)(\mathbb{I} \otimes C)\left[\prod_{t_2}(\mathbb{I} \otimes G^\dagger) S^\dagger (\mathbb{I} \otimes G^\dagger)(\mathbb{I} \otimes C)\right](\mathbb{I} \otimes G^\dagger)(\mathbb{I} \otimes G^\dagger) S(\mathbb{I} \otimes G) |\psi_{t_1}\rangle$$

$$= \left(-e^{i(\phi_1+\phi_2)}\right)^{t_2+1} (\mathbb{I} \otimes C^\dagger)(\mathbb{I} \otimes G)\left[\prod_{t_2+1}(\mathbb{I} \otimes C^\dagger) S^\dagger\right] |\psi_{t_1}\rangle$$

$$= \left(-e^{i(\phi_1+\phi_2)}\right)^{t_2+1} (\mathbb{I} \otimes D)\left(U^\dagger\right)^{t_2+1} (U)^{t_1} |\psi_0\rangle$$